\documentclass[12pt]{article}
\usepackage{amsmath,amssymb}
\usepackage{epsfig,euscript,array}
\usepackage{amssymb}
\usepackage{graphics}
\usepackage{graphicx}

\def\b{\beta}
\def\g{\gamma}

\def\d{\delta}

\def\beq{\begin{equation}}
\def\eeq{\end{equation}}
\def\beqn{\begin{eqnarray}}
\def\eeqn{\end{eqnarray}}
\def\ba{\begin{eqnarray}}
\def\ea{\end{eqnarray}}

\def\m{{\tt -}}

\def\GeV{\,{\rm GeV}}

\def\l{\langle}

\def\xprim2bar{\overline{x}^{\prime\prime}}

\def\beq{\begin{equation}}
\def\eeq{\end{equation}}

\newcommand{\beqa}{\begin{eqnarray}}
\newcommand{\eeqa}{\end{eqnarray}}

   \let\b=\beta   \let\g=\gamma   \let\d=\delta
         
      \let\l=\lambda  \let\m=\mu

\newcommand{\ov}{\overline v}

\newcommand{\ove}{\overline v_{05}}

   \let\b=\beta   \let\g=\gamma   \let\d=\delta
         
      \let\l=\lambda  \let\m=\mu

\newcommand{\be}{\begin{equation}}
\newcommand{\ee}{\end{equation}}
\newcommand{\bea}{\begin{eqnarray}}
\newcommand{\eea}{\end{eqnarray}}

\newcommand{\eq}[1]{Eq.~(\ref{#1})}
\newcommand{\fig}[1]{Fig.~\ref{#1}}
\newcommand{\tab}[1]{Table~\ref{#1}}

\def\A5{(A_5)_{\rm lat}}
\def\thintablerule{\hrule height0.4pt}
\begin{document}


\vskip 1.0cm
\centerline{\LARGE Higgs mechanism near the $5d$ bulk phase transition}

\vskip 2 cm
\centerline{\large Nikos Irges$^1$, Francesco Knechtli$^2$ and Kyoko Yoneyama$^2$}
\vskip1ex
\vskip.5cm
\centerline{\it 1. Department of Physics}
\centerline{\it National Technical University of Athens}
\centerline{\it Zografou Campus, GR-15780 Athens, Greece}
\vskip .4cm
\centerline{\it 2. Department of Physics, Bergische Universit{\"a}t Wuppertal}
\centerline{\it Gaussstr. 20, D-42119 Wuppertal, Germany}
\begin{center}
{\it e-mail: irges@mail.ntua.gr, knechtli@physik.uni-wuppertal.de,\\
yoneyama@physik.uni-wuppertal.de}
\end{center}
\vskip 1.5 true cm
\thintablerule
\vskip 2.0ex
\leftline{\bf Abstract}
\vskip 1.0ex\noindent

We present a non-perturbative model of Gauge-Higgs Unification.
We consider a five-dimensional pure $SU(2)$ gauge theory with orbifold boundary conditions along the fifth dimension, such
that the symmetry is reduced to $U(1)$ at the fixed points of the orbifold action.
The spectrum on the four-dimensional boundary hyperplanes includes, apart from the $U(1)$ gauge boson, also a 
complex scalar, interpreted as a simplified version of the Standard Model Higgs field.
The gauge theory is defined on a Euclidean lattice which is anisotropic in the extra dimension. 
Using the boundary Wilson Loop and the observable that represents the scalar
and in the context of an expansion in fluctuations around a Mean-Field background, we show that
a) near the bulk phase transition the model tends to reduce dimensionally to a four-dimensional gauge-scalar theory,
b) the boundary $U(1)$ gauge symmetry breaks spontaneously due to the broken translational invariance along the fifth dimension,
c) it is possible to construct renormalized trajectories on the phase diagram along which the Higgs mass is constant
as the lattice spacing is varied, 
d) by taking a continuum limit in the regime where the anisotropy parameter is small, it is possible to predict the existence of a 
$Z'$ state with a mass around 1 TeV.

\vskip 2.0ex
\thintablerule

\vskip-0.2cm
\newpage

\section{Introduction}

In the Standard Model (SM) the Higgs field $H$ is introduced as a fundamental scalar and inserted in the 
classical Lagrangean via the most general potential of engineering dimension four,
consistent with the field's quantum numbers:
\be
V_H = -\m^2H^\dagger H + \l (H^\dagger H)^2\, .
\ee
The relative sign between the two terms is not fixed by any symmetry and is an external assumption.
The Higgs mechanism then automatically proceeds
by the field developing a vacuum expectation value (vev) $<H> = v/\sqrt{2}$ which, upon minimization of the potential, turns out to be non-zero
and satisfying $v=\m/\sqrt{2}$ at the classical level. At the same order, the Higgs mass is $m_H = v \sqrt{2\l}$ and the neutral gauge boson mass
$m_Z = 1/2 v g$ with $g$ a coupling derived from the gauge couplings of the group factors that contribute to the mass.
Thus, in its minimal version, the Higgs mechanism is described by a gauge coupling $g$, a dimensionless quartic coupling 
$\l$, a dimensionful mass parameter $\m$ and the vev $v$. External input is also the potential itself, in the sense described above.
Fluctuations around this non-trivial vacuum define the SM in its
spontaneously broken phase, a state of matters that seems to be consistent even with the most recent LHC data.
The origin of the ingredients that conspire to make the mechanism work is however unknown.
Perhaps the most convincing clue that this cannot be the end of our formulation of the low energy end of 
high energy elementary particle physics is the quantum response of the fluctuations. There is a quadratic dependence of the
scalar mass on the cut-off, believed to render the theory unnatural for a light Higgs particle. 
Historically the dominant solution to this puzzle has been supersymmetry. 
Here we propose an alternative scenario where the mechanism develops dynamically, described by three 
(in infinite lattice volume actually only by two) dimensionless
parameters, without introducing an explicit potential or a vev. Moreover, we demonstrate that in our 
proposed scheme the mass of the Higgs particle is insensitive to the cut-off along renormalized trajectories, without supersymmetry.

The general context is that of "Gauge-Higgs Unification" (GHU) \cite{Manton:1979kb,Hosotani:1983vn}, where the Higgs field originates from the components of
a higher dimensional gauge field along the extra dimension(s). Since we would like to have a control of the theory
at the quantum level, we will exclude from our discussion warped and curved space-times.
Instead, our starting point is a five-dimensional $SU(2)$ gauge theory compactified on the $S^1/Z_2$ orbifold,
possibly the simplest prototype GHU model where the dynamical Higgs mechanism can be studied.
The boundary conditions introduce four-dimensional boundaries at the fixed points of the orbifold,
where the gauge symmetry is reduced to $U(1)$. Subsequently, the boundary symmetry can, in principle, break
spontaneously, generating a massive $Z$ boson and a massive Higgs.
Originally, this model was studied in the perturbative regime where the first peculiar 
deviation from what we see in the SM was recognized to be
the fact that fermions are necessary to trigger spontaneous symmetry breaking (SSB),
tying the existence of SSB to the fermionic content \cite{Hosotani:1983vn,Kubo:2001zc}.
Another obscure fact that adds to the above is that even if one accepts this as a necessary feature of GHU models,
it seems to be hard to have a Higgs heavier than the neutral $Z$ gauge boson,
without considering values for the parameters involved somewhat unnatural.
Also, in the perturbative approach, even though the scalar potential is a dynamical quantum (Coleman-Weinberg-Hosotani (CWH)) effect,
a vev must be inserted by hand, like in the SM. Finally, the non-renormalizability 
of the gauge theory often brings doubts about certain attractive conclusions 
drawn from perturbative loop calculations, such as the finiteness of the Higgs mass \cite{vonGersdorff:2002as,Cheng:2002iz}.
From our point of view, it is this non-renormalizability along with the perturbative triviality of the five-dimensional gauge coupling 
(perhaps not an unrelated property to the former) that may be the root of these problems. 

For the above reasons, we have started a non-perturbative investigation of these simple 
orbifold gauge theories \cite{Irges:2004gy}.\footnote{Recent Monte Carlo investigations of the 
periodic theory include  \cite{Ejiri:2000fc,deForcrand:2010be,Farakos:2010ie,Knechtli:2011gq,DelDebbio:2012mr}.}
At an early stage, an exploratory lattice Monte Carlo (MC) study of the $SU(2)$ orbifold theory was performed that revealed that SSB is present already
in the pure gauge system, signaled by a massive $Z$ boson \cite{Irges:2006zf,Irges:2006hg}.
At the same time however also the practical difficulty of a systematic MC study and the necessity for a non-perturbative
analytic approach became obvious. Recently such a formalism was developed \cite{Irges:2009bi,Irges:2012ih}, consisting of an
expansion of the path integral in fluctuations around a Mean-Field (MF) background.
Furthermore, introducing an anisotropy along the fifth dimension proved to be fruitful. 
There is increasing confidence by now that this MF expansion is a faithful representation of the 
non-perturbative system in five dimensions.
%
\begin{figure}[!t]
\begin{minipage}{.49\textwidth}
\centerline{\epsfig{file=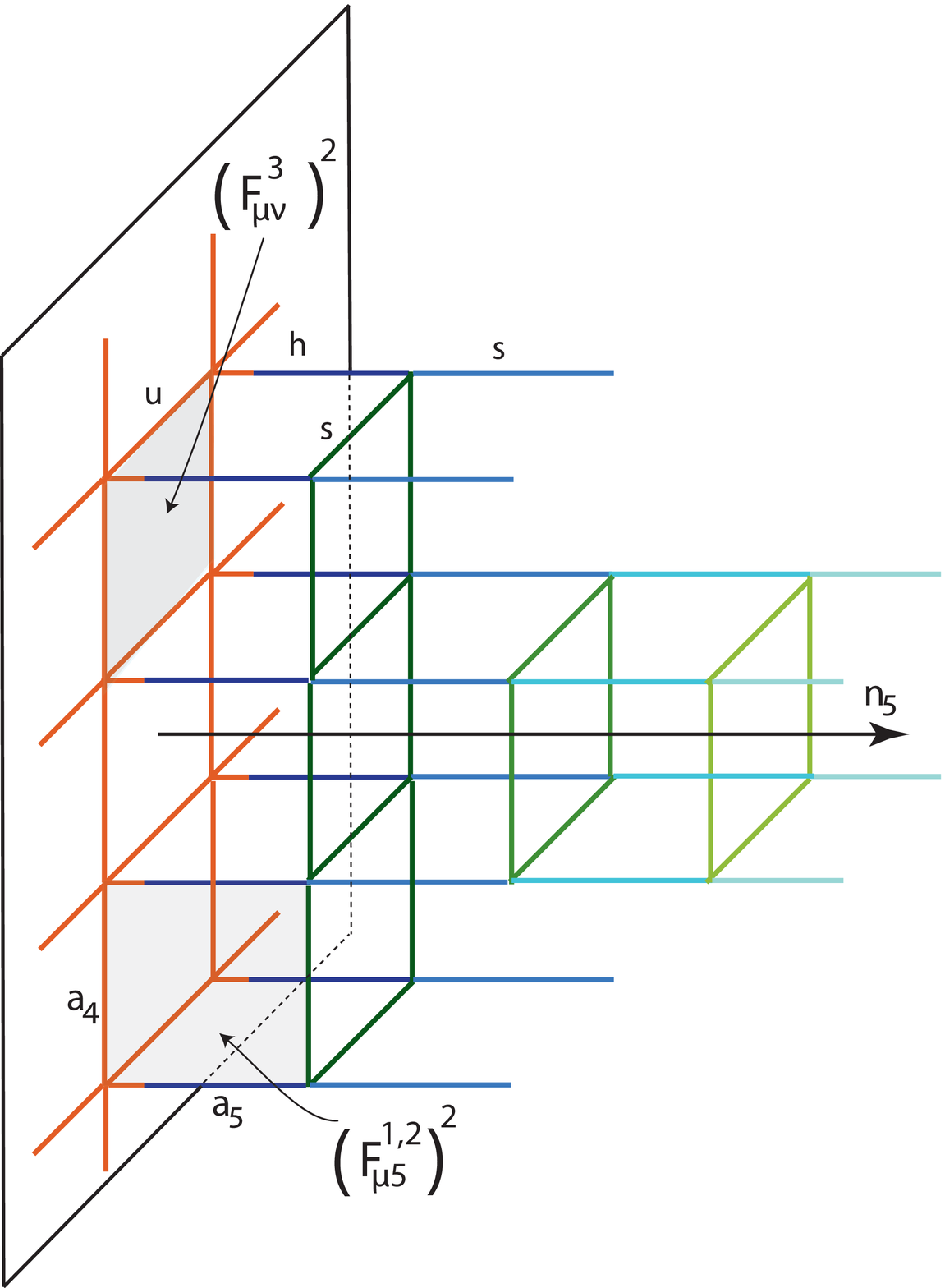,width=7cm}}
\end{minipage}
\begin{minipage}{.49\textwidth}
\centerline{\epsfig{file=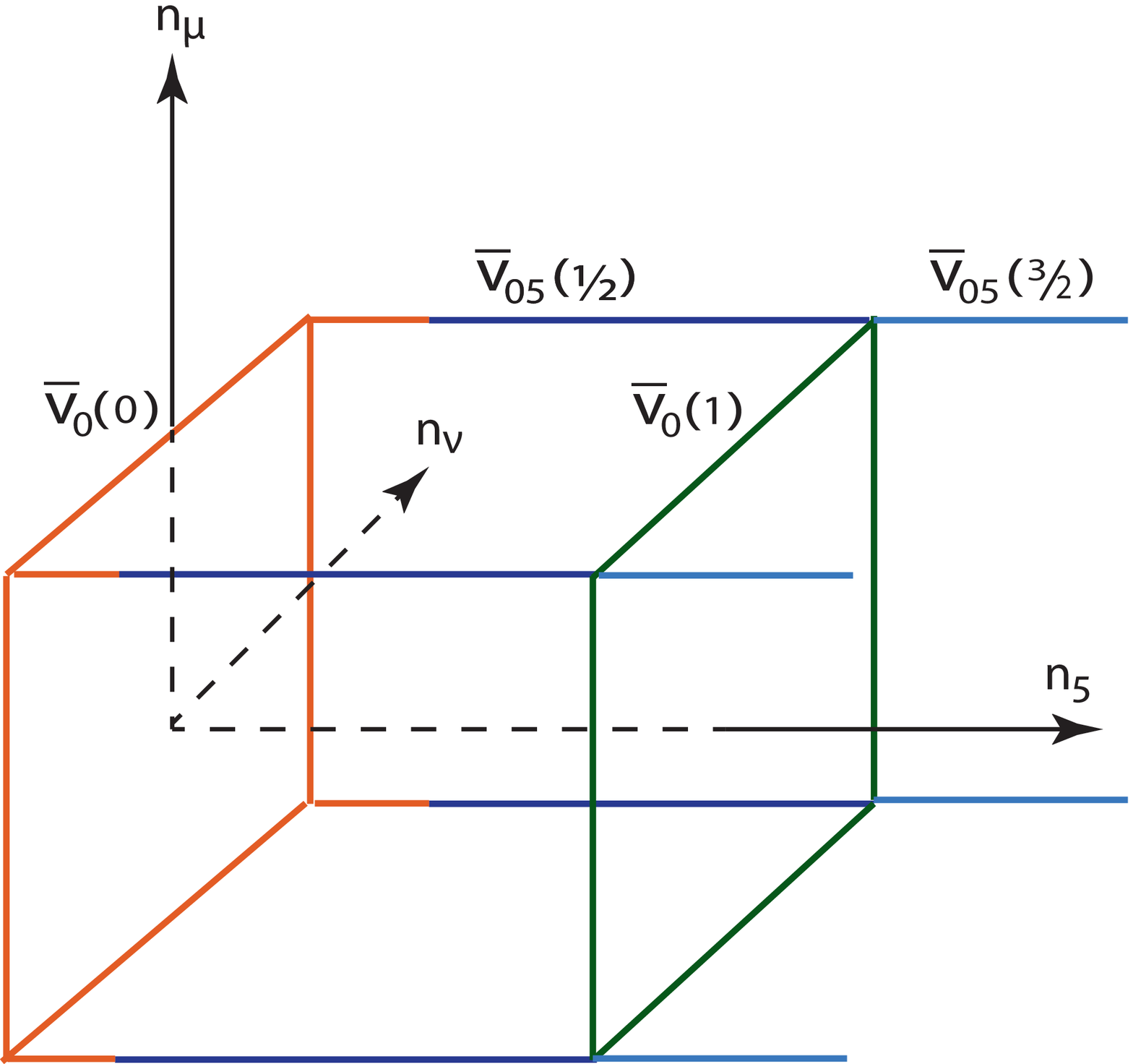,width=7cm}}
\end{minipage}
\caption{The lattice orbifold. The Mean-Field background depends on the
extra-dimensional coordinate $n_5$ in a way that induces the
spontaneous breaking of the gauge symmetry. Lighter shades of a given color indicate
different MF background values ${\overline v}_0(n_5+k/2)$, $k=0,1,2,3,\cdots$ of the links.
\label{f_ssb}}
\end{figure}
%
If the five-dimensional SU(2) gauge theory has periodic boundary conditions along
the extra dimension, dimensional reduction to four dimensions (if it occurs) drives the system to a 
Georgi--Glashow model (i.e. a $4d$ $SU(2)$ gauge theory coupled to an adjoint scalar) from the lower dimensional point of view. 
Our Mean-Field analysis suggests that with these boundary conditions SSB does not occur.
If instead we consider orbifold boundary conditions, due to the breaking of
the $SU(2)$ gauge symmetry to $U(1)$ at the boundaries, dimensional reduction takes us to
a four-dimensional Abelian-Higgs system. Our Mean-Field expansion method indicates that
SSB is realized in this case  \cite{Irges:2012ih}. In fact, the static potential along the boundaries is of a $4d$ Yukawa type and
the (smallest) Yukawa mass corresponds to the mass of the $Z$ gauge boson,
confirming the earlier Monte Carlo simulations of the orbifold model. 

In order to understand the origin of SSB, we have to take a closer look at the structure of our construction.
On the left of \fig{f_ssb} we show a schematic picture of the orbifold lattice we have in mind. Its precise definition
has been given elsewhere \cite{Irges:2004gy}. The four-dimensional boundaries at the two fixed points of the orbifold
action are covered by $U(1)$ links ($u$-links). All other links are $SU(2)$ links ($s$-links), except from those
lying along the fifth dimension and whose one end touches the boundary ($h$-links).
These have one ''end'' transforming as U(1) and the other as SU(2).
This is the proper orbifold lattice, invariant under the gauge and orbifold actions.
In fact, all correlators representing physical observables like the Wilson Loop
and the Higgs are also gauge invariant and invariant under the orbifold action \cite{Irges:2004gy}.
As a result, the breaking of translational invariance along the extra dimension is spontaneous.
In a 1-loop CWH computation this breaking is encoded merely in a modification of the pre-factor of the potential, 
leading to the conclusion of no SSB, as on the torus. 
In the present approach on the other hand already the Mean-Field background contains the necessary information. 
The Mean-Field equations for the background produce a non-trivial profile for it along the extra dimension \cite{Knechtli:2005dw,Irges:2012ih},
schematically represented on \fig{f_ssb}, where different shades of a given
color represent different MF background values of the links. On the right of the figure we have
explicitly indicated a few of those values.
The non-trivial profile is of course a direct consequence of the spontaneously broken translational invariance,
allowing us to call sometimes the MF background (together with the vibrations of the lattice above it) 
as a "phonon" in a condensed matter language.
For similar reasons, the spontaneously broken translational invariance is felt by 
the system without the need for any background, in a full Monte Carlo study \cite{Irges:2006zf,Irges:2006hg}.

Finally in regards to the connection of our MF approach to perturbation theory, we mention that each order of this expansion
is believed to be equivalent to the summation of an infinite number of perturbative Feynman diagrams \cite{Drouffe:1983fv}.
Practically, by trivializing the background and taking the lattice coupling to infinity, we can reach the perturbative regime
and reproduce the corresponding results.

\section{The Higgs mechanism as a phonon trigger effect}

Our anisotropic lattice has $L$ spatial points, $T$ time-like points (with lattice spacing $a_4$) and $N_5+1$ points along the fifth dimension
(with lattice spacing $a_5$). 
$L$ will be removed by increasing it enough so that physics does not depend on it.
Time will be used to extract the masses of the ground and first excited states.
In short, the five-dimensional lattice coupling $\b=4a_4/g_5^2$ ($g_5$ is the dimensionful 5d gauge coupling), the anisotropy parameter $\g=a_4/a_5$ (at the classical level) and $N_5$
are the three dimensionless parameters that parametrize our model.
The action is the five-dimensional Wilson plaquette action, anisotropic in the fifth dimension.

The quantities that will be used in our analysis are the Wilson Loop 
on the boundary at the origin and the Higgs observable.
In \cite{Irges:2012ih} we computed them to first non-trivial order in the fluctuations around the MF
background and we will not repeat the calculations here. We will instead compute these observables numerically
and extract the static potential from the former and the Higgs mass from the latter. 
The specific quantities that we are interested in are
\be
F_1=m_H R\, ,
\ee
the Higgs mass in units of the inverse size of the extra dimension $1/R$, the ratio
\be
\rho_{HZ} = \frac{m_H}{m_Z} \label{rho}
\ee
of the Higgs to $Z$ mass, with the $Z$ mass extracted from a fit of the static potential to the form
($b$ is a constant and $r$ is the spatial length of the Wilson Loop)
\be
V(r) = -b \frac{e^{-m_Z r}}{r} + {\rm const.} \label{static}
\ee
and 
\be
\rho_{HZ'} = \frac{m_H}{m_{Z'}}\, ,\label{rhoprime}
\ee
the ratio of the Higgs mass to the mass of the first excited vector boson state. This is usually called a $Z'$,
and it can also be derived straightforwardly from a fit of the static potential. 
Clearly, such a fit makes sense if the spectrum can be interpreted as an effective four-dimensional theory,
which by itself is not a precise enough definition of a satisfactory dimensional reduction.
Our more constrained criteria for dimensional reduction are therefore that
\begin{itemize}
\item the fit to \eq{static} is possible with $m_Z\ne 0$. 
This ensures that there is SSB, signaled by the presence of the massive $U(1)$ gauge boson.
Otherwise the gauge boson is massless and only a Coulomb fit is possible.
\item the quantities $M_H=a_4 m_H$ and $M_Z=a_4 m_Z$ are $<1$. 
This ensures that we are in a regime of the phase diagram where the lattice spacing does not dominate the observables.
\item we have $m_HR<1$ and $\rho_{HZ}>1$. These two conditions ensure that the Higgs and the $Z$ mass are lighter
than the Kaluza-Klein scale $1/R$ on one hand and that the Higgs is heavier than the $Z$ on the other, a desirable situation
from the phenomenological point of view. In fact, we will target the value
\be
\rho_{HZ} = 1.38\, ,\label{SMrho}
\ee
which is (approximately) the currently favored value of the analogous quantity in the SM, based on recent LHC data \cite{ATLAS:2012gk,CMS:2012gu}.
\end{itemize}
In summary, we have the observables $F_1$, $\rho_{HZ}$ and $\rho_{HZ'}$, all three depending on the three
dimensionless parameters $\b$, $\g$ and $N_5$.
Our method then is to fix $F_1$ to a given value and keep $\rho_{HZ}$ fixed to the value \eq{SMrho}. 
This leaves $\rho_{HZ'}$ be a function of one parameter which we choose to be $N_5$ and
by doing so we obtain a value for the mass of the $Z'$ for each $N_5$. 
We call such a trajectory on the phase diagram that also fulfills our three conditions for dimensional reduction,
a Line of Constant Physics (LCP).
For illustrative purposes we will also give dimensionful values for the masses by 
inserting in the ratios the SM value for $m_Z$.

Eventually we would like to understand the physical meaning of the $N_5\to \infty$ limit 
and for that we have to describe the structure of the phase diagram. 
In the MF expansion the phase diagram can be plotted already at the level of the background.
Even though one could consider corrections due to fluctuations, we will stay at the lowest order,
because the corrections can be seen to be small. 
The background value of the gauge link variables on the anisotropic lattice are denoted 
as $\ov_0(n_5)$ along the $\m=0,1,2,3$ directions and as $\ove(n_5+1/2)$ along the fifth dimension,
with $n_5$ denoting the corresponding integer coordinate.
The phases of the system can be defined from the phonon profile as follows (the statements hold $\forall n_5$):
\begin{itemize}
\item Confined phase: $\ov_0(n_5), \ove(n_5+1/2) = 0$.
\item Layered phase: $\ov_0(n_5) \ne 0$, $\ove(n_5+1/2) = 0$.
\item Deconfined phase: $\ov_0(n_5), \ove(n_5+1/2) \ne 0$.
\end{itemize}
According to the MF method, the boundary between the Deconfined phase and the Confined and Layered phases 
has a different order depending on $\b$ and $\g$. Tuning $\b$ so that one is in the Deconfined phase and always near the phase boundary,
one finds that for $\g$ larger than a value that is slightly less than 1, the phase transition is of first order,
while below that value it turns into second order. We emphasize that the phase transitions defined in this way
are always {\it bulk}  phase transitions, a fact that has been extensively verified on the periodic lattice \cite{Knechtli:2011gq}.
Even though it is much harder to verify via Monte Carlo simulations the change of the order of the phase transition at small $\g$ \cite{Farakos:2010ie},
we will assume here that the order of the phase transitions that the MF predicts is always correct.

When the bulk phase transition is of first order, the four-dimensional lattice spacing
$a_4$ remains finite no matter how closely the phase boundary is approached. 
Then, since ($l$ is the physical four-dimensional volume)
\be
L = \frac{l}{a_4}\, ,\hskip 2cm N_5 = \frac{\pi R}{a_4}\g\, ,
\ee
in the $L=N_5=\infty$ limit the physical size of the system goes to infinity at a finite lattice spacing.
When the phase transition is second order, one expects instead that the lattice spacing goes to zero at a finite
physical volume. There is an important physical difference between the two cases.
In the first case the low energy theory is an effective theory that must be defined with a finite cut-off. 
The existence of the LCP nevertheless guarantees that a sensible (in the quantum sense) and predictive
effective theory exists.
In the second case SSB would persist in the continuum limit and there is no need for a cut-off in the effective action.
This would render the theory non-perturbatively renormalizable. 

With or without continuum limit, we will show that in the vicinity of the phase boundary 
we have a gauge-scalar system without a hierarchy problem and with a dynamical Higgs mechanism
that may be described as a phonon trigger effect:
as in crystals where the formation of Cooper pairs takes place and whose interaction with the phonon leads to superconductivity,
here the Polyakov Loop with quantum numbers appropriate for the Higgs boson
(of charge 2 under the U(1), see \cite{Irges:2006hg}) interacts with the MF background
triggering the spontaneous breaking of the gauge symmetry.
Like on the periodic lattice, our motivation to believe that the lattice spacing decreases as the phase transition is approached is that
both $M_H$ and $M_Z$ decrease. In fact, the only place where one can have $M_H,M_Z < 1$ is reasonably near the phase transition.
In the small $\g$ regime one can have values pretty much as small as desired.
In the large $\g$ regime on the other hand there seems to be a barrier below which the masses can not be decreased anymore.
Therefore it is not a surprise that if it is at all possible to construct an LCP as defined above, it will surely be a line near the phase boundary.
\begin{table}
\begin{center}
\begin{tabular}{c|c|c|c}
$F_1$ & $N_5$ & $\gamma^*$ & $\beta^*$ \\
\hline
0.61 & 12 & 0.5460(33) & 1.343501425\\
&14 & 0.5320(10) & 1.34442190 \\
&16 & 0.5228(7)  & 1.34664820 \\
&20 & 0.5028(18) & 1.35582290 \\
&24 & 0.4844(32) & 1.36940695 \\
\hline
0.20 & 6  & 0.5113(15) & 1.351160631
\end{tabular}
\end{center}
\caption{Bare parameters of the LCP defined by
$\rho_{HZ} = m_H/m_Z = 1.38$ and $F_1 = m_H R = 0.61$,
together with one point for a LCP with $\rho_{HZ} = 1.38$ and $F_1 = 0.20$.
The lattice gauge couplings $\beta^*$ correspond to the central values $\gamma^*$
and are computed for future reference.}
\label{t_LCPparams}
\end{table}
%
\begin{figure}\centering
  \resizebox{12cm}{!}{\includegraphics[angle=0]{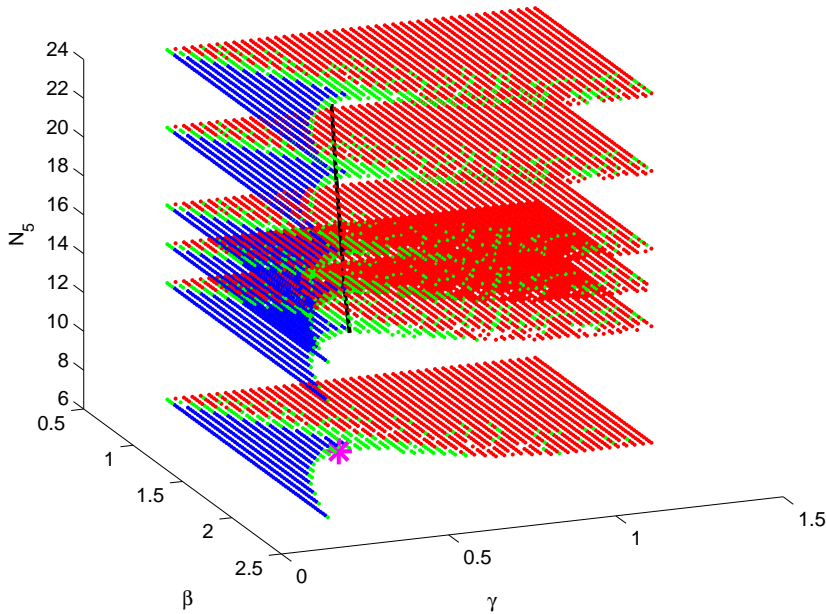}}
  \caption{LCP (black line) defined in \eq{lcp1} 
  near the (tricritical point of the) bulk phase transition. 
  Red: Confined phase. Blue: Layered phase. White: Deconfined phase.
  The magenta point (star) is on a different LCP with $F_1=0.2$, $\rho_{HZ}=1.38$.}
  \label{LCPphasediagram}
\end{figure}
%

\section{Lines of Constant Physics and the $Z'$}

The first LCP we construct is one where
\be
F_1 = m_H R = 0.61\, , \hskip .5cm \rho_{HZ} = 1.38\, \label{lcp1}
\ee
are kept fixed. Along this LCP, we compute $\rho_{HZ'}$ for $N_5=12, 14, 16, 20, 24$. 
On \fig{LCPphasediagram} we plot the corresponding points interpolated by a black line
on the phase diagram, which are listed in \tab{t_LCPparams}.
As discussed, it is a line near the phase boundary and in particular in the small $\g$ 
regime where the phase transition is of second order according to the MF. Thus we can 
attempt to take the continuum limit. Note that this is also the regime where from both 
MF \cite{Irges:2009qp} and MC \cite{Knechtli:2011gq}
studies of the periodic lattice it has been seen that gauge fields are localized
on four-dimensional hyperplanes. Evidently this is an effect independent of boundary conditions.

\begin{figure}[!t]
\begin{minipage}{.49\textwidth}
\centerline{\epsfig{file=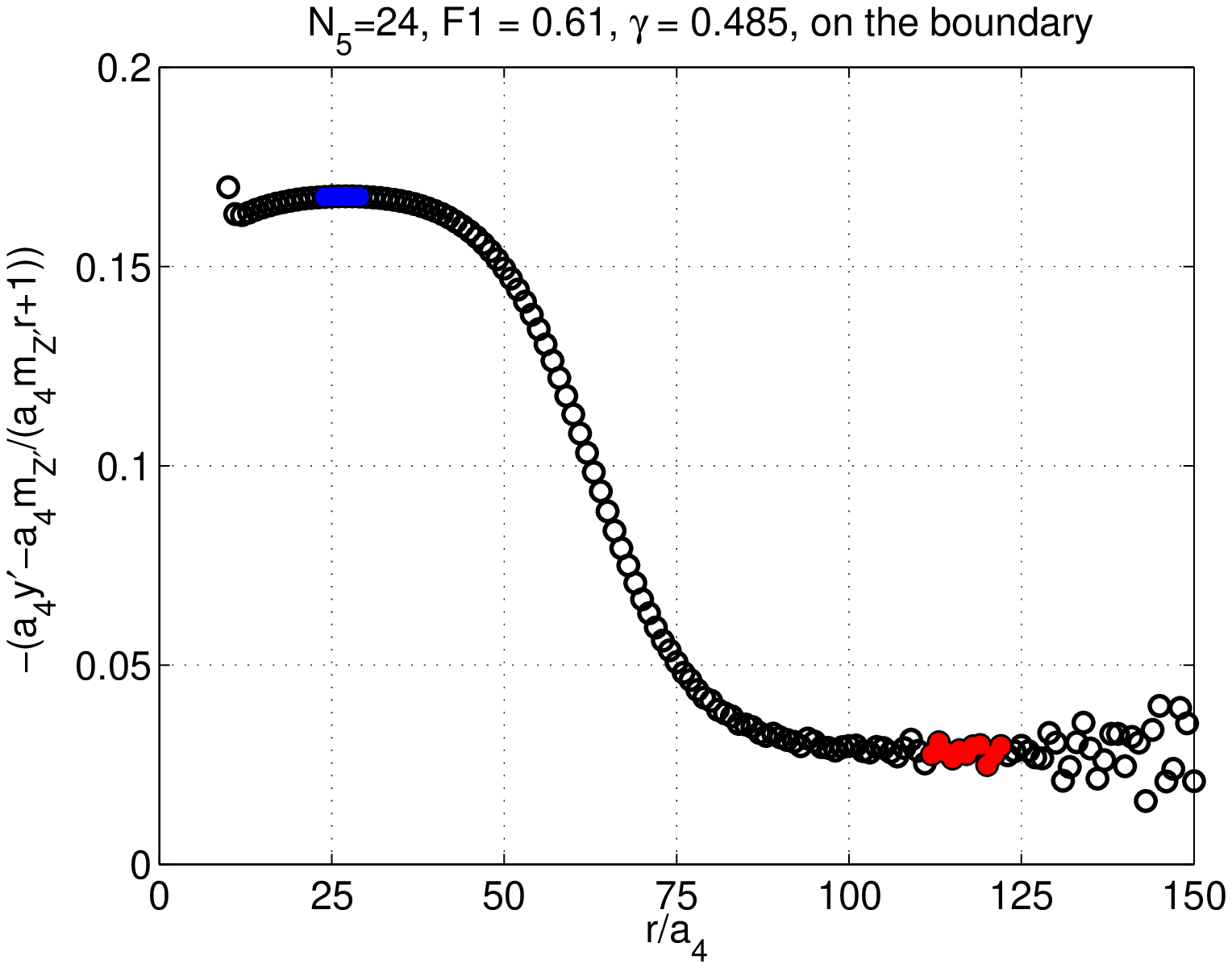,width=8cm}}
\end{minipage}
\begin{minipage}{.49\textwidth}
\centerline{\epsfig{file=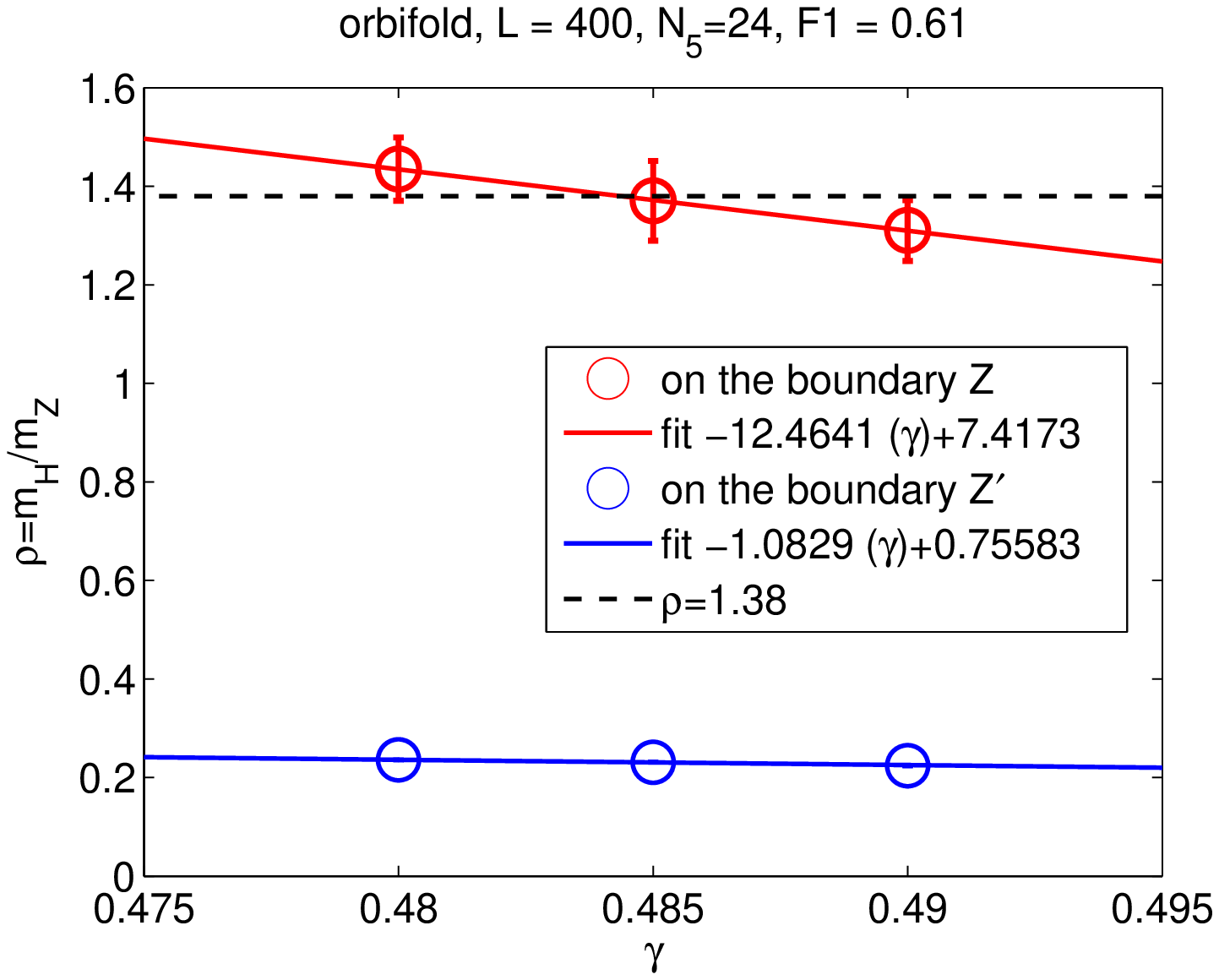,width=8cm}}
\end{minipage}
\caption{Left plot: plateaus of the quantity defined in \eq{yprime}
corresponding to the $Z$ (red points) and $Z'$ (blue points) masses.
Right plot: the $\rho_{HZ}$ data (upper red circles) are lineraly interpolated 
(red line) to the value of $\gamma$ corresponding to $\rho_{HZ}=1.38$ 
(marked by the dashed horizontal line). 
The lower blue circles show the data for $\rho_{HZ'}$ with a linear fit
(blue line).
\label{f_yprime_rho}}
\end{figure}

For each value of $N_5$, we compute the $Z$ and $Z'$ masses for various values
of the parameter $\gamma$. The third parameter $\beta=\beta(\gamma,N_5)$ is set by
requiring that $F_1$ has the desired value $0.61$.
The gauge boson masses are extracted by identifying them as Yukawa masses
describing the static potential $V(r)$ along the boundary.
From the static force $F={\rm d}\,V(r)/{\rm d}\,r$ we compute the quantity
\be
y'(r)=[y(r+a_4)-y(r-a_4)]/(2a_4) \,,\qquad y(r) = \ln(r^2F(r)) \,. \label{yprime}
\ee
A Yukawa mass is identified as a plateau in the quantity $-y'(r)$ as a function of $r$.
As it is shown on the left plot in \fig{f_yprime_rho} for $N_5=24$ and $\gamma=0.485$,
typically there are two plateaus, the smaller (red points) we take to define $M_Z$ and the 
larger (blue points) $M_{Z'}$.
The ranges of $r$ values defining the plateaus are taken around the minima of the derivative 
of $-y'(r)$. 
The masses are the average and the errors are the standard deviation of the plateau points.
The value of $M_Z$ is improved by iteratively solving
$M_Z = -\left[a_4y'(r) - M_Z/(M_Zr/a_4+1)\right]$.
The value of $L$ should be large enough to clearly identify the plateaus and we set
$L=400$ for all $N_5$ values.
%
%

In the next step, at fixed $N_5$, we compute $\rho_{HZ}$ as a function of $\gamma$
from the known values of $M_Z$ and $M_H$. The data can be very well fitted by a straight
line, as is demonstrated for $N_5=24$ by the upper red points and the red line 
on the right plot of \fig{f_yprime_rho}. From the linear
fit we compute the value $\gamma^*(N_5)$ which gives the desired value $\rho_{HZ}=1.38$.
The error on $\gamma^*$ takes the correlation of the fit parameters into account. For the
data shown in \fig{f_yprime_rho} we get $\gamma^*(24)=0.4844(32)$. A summary of the
LCP parameters for all $N_5$ values is given in \tab{t_LCPparams}.
%
\begin{figure}\centering
  \resizebox{12cm}{!}{\includegraphics[angle=0]{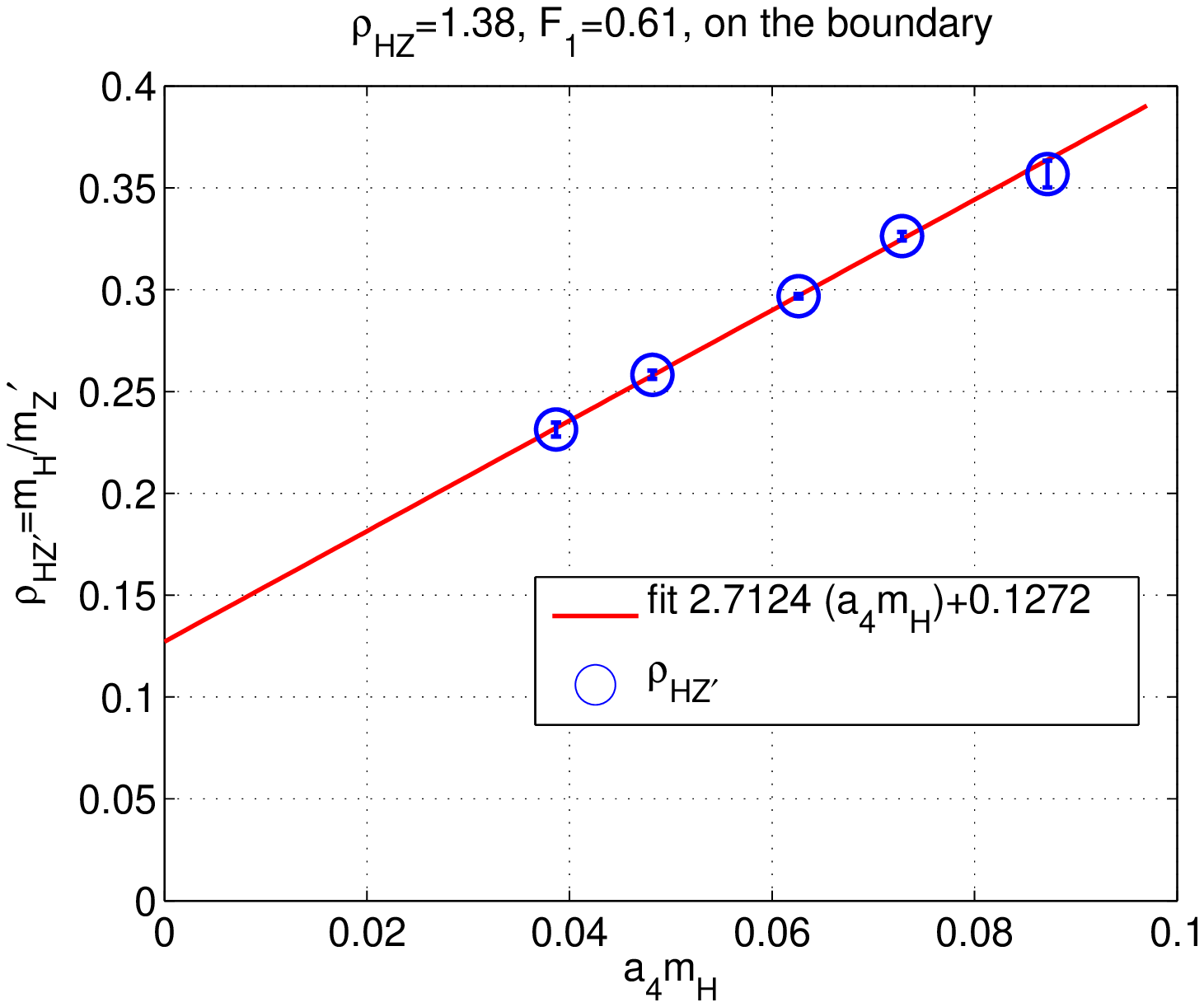}}
  \caption{Extrapolation of LCP in \eq{lcp1} to $a_4m_H\to 0$.}
  \label{LCPZprime}
\end{figure}
%

Finally we compute, for each value of $N_5$, the ratio $\rho_{HZ'}$ when
$\gamma$ is set to the value $\gamma^*$. This is done by fitting linearly in
$\gamma$ the data of $\rho_{HZ'}$ as shown by the lower blue points and blue
line on the right plot of \fig{f_yprime_rho}. We take the fit result evaluated at $\gamma^*(N_5)$
and augment the fit error by adding the slope of the fit multiplied by the
uncertainty of $\gamma^*(N_5)$. For the
data shown in \fig{f_yprime_rho} we get $\rho_{HZ'}(\gamma^*)=0.2313(1)(34)$, where
the dominant error comes from the uncertainty of $\gamma^*(N_5)$.
On \fig{LCPZprime} we plot our data of $\rho_{HZ'}$ against $a_4m_H$ for
$N_5=12, 14, 16, 20, 24$. 
Notice that since $F_1=m_HR=(a_4m_H)N_5/(\gamma^*\pi)$, 
$a_4m_H$ is proportional to $\gamma^*/N_5$ along the LCP and it is $a_4m_H$ which
measures the physical distance to the continuum limit (it is the inverse
correlation length). In principle we could fit the points with a quadratic curve because from the
Symanzik analysis of cut-off effects we expect the 
dominant contributions to come from the dimension 5 boundary operator
\be
\frac{\pi}{4} \left(F^1_{5\m}F^1_{5\m} + F^2_{5\m}F^2_{5\m} \right) \d_{n_5,0} 
\label{Sym1}
\ee
multiplied by one power of the lattice spacing and from the dimension 7 bulk operator
\be
\frac{1}{2g_5^2} \frac{1}{24}\sum_{M,N} {\rm tr} \left \{ F_{MN} \left( D_M^2 + D_N^2\right) F_{MN} \right \} 
\label{Sym2}
\ee
multiplied by two powers of the lattice spacing \cite{Irges:2007qq}. 
In fact, because we are very close to the phase transition, we are in a regime where the effect of
the dimension five boundary operator dominates, thus the linear fit on \fig{LCPZprime}.
The extrapolation to $a_4m_H\to 0$ leaves a
non-zero intercept with the vertical axis which corresponds to
\be
\rho_{HZ'} = 0.1272\, .
\ee
For $m_Z=91.19\GeV$ this implies a $Z'$ of mass $m_{Z'}=989\GeV$ in the continuum limit.
The $\chi^2$ per degree of freedom of the fit is excellent, $0.025/3$.

\begin{figure}[!t]
\begin{minipage}{.49\textwidth}
\centerline{\epsfig{file=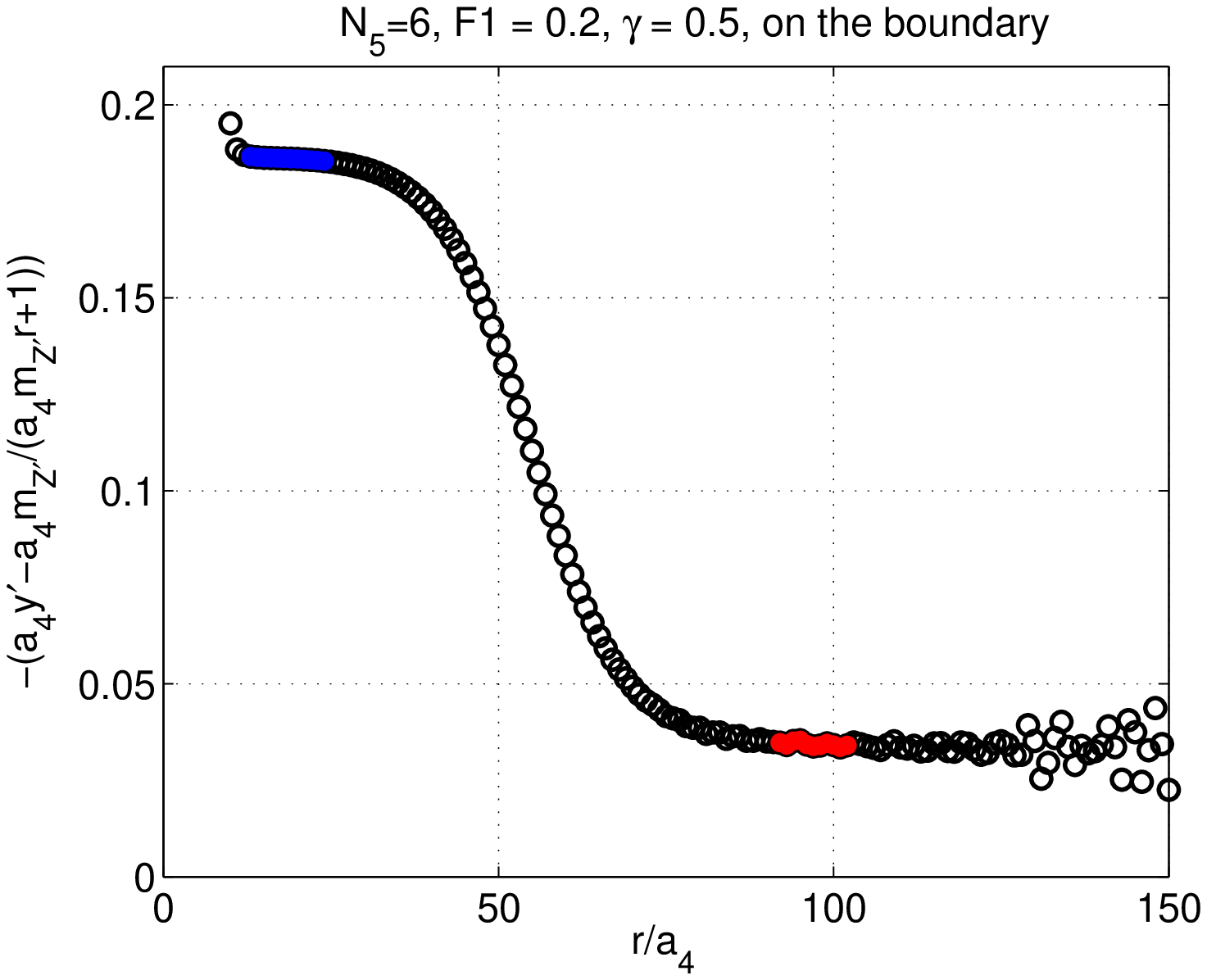,width=8cm}}
\end{minipage}
\begin{minipage}{.49\textwidth}
\centerline{\epsfig{file=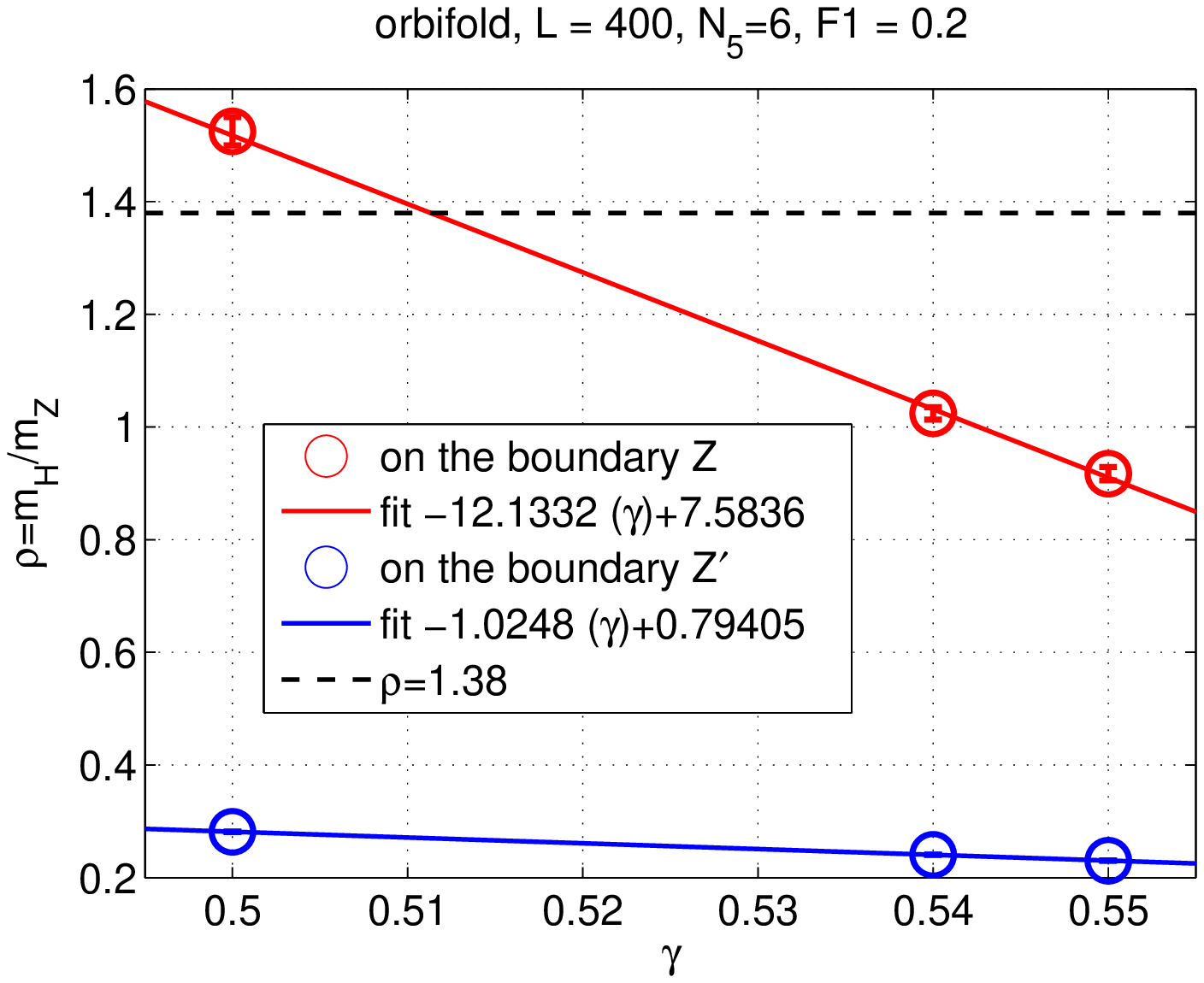,width=8cm}}
\end{minipage}
\caption{Same as in \fig{f_yprime_rho}  but for $F_1=0.20$.
\label{f_yprime_rho_0p2}}
\end{figure}

Before we state our conclusions, it is worth making a comparison between the SM Higgs mechanism and our scheme.
As these are two different theories, there is no well defined way to do this so we can only be qualitative.
The dimensionless 4d gauge coupling $g$ can be thought to be analogous to the dimensionless
5d gauge coupling $\b$. The Higgs field $H$ is introduced in the SM by hand while in the lattice GHU model
its presence is induced by the orbifold boundary conditions, which reduce the adjoint scalar
$A_5^A$ to the complex scalar made from $A_5^{1,2}$. The quartic coupling $\l$ of the SM Higgs sector
is a dimensionless parameter like the anisotropy parameter $\g$. The choice of the negative sign in front of the
$\m^2$ term in the Higgs potential triggers SSB in the SM. In the MF expansion we see SSB once the background is
defined as the point around which the path integral is expanded, a signature of the broken translational invariance
in the extra dimension. 
The presence of the phonon seems to be the crucial fact that triggers the 
subsequent spontaneous breaking of the gauge symmetry, like in superconductors.
In the SM it is likely that the parameter $\rho_{HZ}$ takes the approximate value 1.38, an experimental fact. 
On our lattice orbifold there is a family of $\rho_{HZ}=1.38$ LCP's, each labelled by a different value of $m_HR$ and we have chosen
to plot one of these, the one that corresponds to $F_1=m_HR=0.61$. 
We have checked that there is in principle no obstruction in constructing an LCP with $\rho_{HZ}$=1.38 and a much smaller $F_1$
(we have generated a point for $F_1=0.20$, see \tab{t_LCPparams} and \fig{LCPphasediagram} and \fig{f_yprime_rho_0p2}), but numerically this is slightly more demanding
since the masses in lattice units become very small as a function of $N_5$.

We checked also the region where $\gamma$ is one or larger and found that
an LCP with $\rho_{HZ}=1.38$ is possible only in the small gamma regime.
In the regime $\gamma\ge1$ dimensional reduction can occur only through compactification and
therefore we choose $F_1$ between $0.10$ and $0.20$.
For $\gamma$ around one we get $\rho_{HZ}$ values which are smaller than but close to one,
in agreement with the results from Monte Carlo simulations at $\gamma=1$ 
\cite{Irges:2006zf,Irges:2006hg,Irges:2007qq,Knechtli:2007ea}.
The $\rho_{HZ'}$ values are $\approx0.090$.
For $\gamma=4$ we see only one plateau for the quantity defined in \eq{yprime}, which we
interpret as the $Z$ mass. The $\rho_{HZ}$ values are again consistent with one.
The fact that we do not see a second plateau for the $Z'$ mass
may be reasonable since we are in the compact phase and it could 
be that the $Z'$ state is too heavy.
%
%

\section{Conclusions}

We presented a numerical analysis of observables computed in an analytical expansion in fluctuations around 
the Mean-Field background on the anisotropic lattice orbifold, defined and developed in previous publications.
We showed that non-perturbatively spontaneous symmetry breaking is a property of the pure gauge system and
we were able to draw lines on the phase diagram along which the ratio of the Higgs over the $Z$ 
boson mass $\rho_{HZ}$ remains finite. Furthermore, by taking the continuum limit near the bulk phase transition and 
at values of the anisotropy parameter around $\g \simeq 0.5$, we demonstrated that the first excited state
in the vector boson sector is light, with a mass in the TeV regime. The Standard Model value of $\rho_{HZ}=1.38$
for which this Line of Constant Physics was drawn, could be reached only in this, small $\gamma$ regime.
Even though we used a toy $SU(2)$ model, we believe that these generic properties will persist in more realistic cases
where the bulk gauge symmetry may be for example  $SU(3)\times SU(3)$ or $SU(5)$.
Similarly, we expect that the presence of fermions will not alter the observed qualitative behavior.
We plan to perform these generalizations in the future.
The scheme of the Mean-Field expansion seems to be a good (semi)analytical description of the
non-perturbative system in five dimensions thus it should be considered at least as a valuable complementary
tool in the study of Gauge-Higgs Unification. For among others, it provides us with a guide to a
Monte Carlo study, pointing to the regime on the phase diagram where one should perhaps focus.
Especially if the absence of low energy supersymmetry is experimentally confirmed, our results here
may have given us a hint for an alternative solution to the Higgs hierarchy problem and for a possible
dynamical, non-perturbative origin of the Higgs mechanism.

{\bf Acknowledgments.} We thank Andreas Kl{\"u}mper for discussions. 
K. Y. is supported by the Marie Curie Initial Training Network STRONGnet.
STRONGnet is funded by the European Union under Grant Agreement number 238353 
(ITN STRONGnet). N. I. thanks the Alexander von Humboldt Foundation for support.
N. I. was partially supported by the NTUA research program PEBE 2010, 65184800.

\bibliography{orbifold}     
\bibliographystyle{h-elsevier}   
\end{document}